\documentclass[11pt]{article}
\usepackage[cp866]{inputenc}
\usepackage[russian,english]{babel}
\usepackage{epsf}

\begin{document}

\title{ \bf Cooling of Neutron Stars: Two Types
of Triplet Neutron Pairing}

\author{ {\bf M. E.\ Gusakov$^1$\footnote{{\it e-mail}:
gusakov@astro.ioffe.rssi.ru} $\,$,
 O. Y. Gnedin$^2$} \\
     {\it $^1$  Ioffe Physical Technical Institute,} \\
     {\it Politekhnicheskaya 26, 194021 St. Petersburg, Russia} \\
{\it $^2$ Space Telescope Science Institute,} \\
{ \it 3700 San Martin Drive, Baltimore, MD 21218, USA} }
\date{${}$}
\maketitle
\def\la{\;\raise0.3ex\hbox{$<$\kern-0.75em\raise-1.1ex\hbox{$\sim$}}\;}
\def\ga{\;\raise0.3ex\hbox{$>$\kern-0.75em\raise-1.1ex\hbox{$\sim$}}\;}
\def\pFn{p_{\raise-0.3ex\hbox{{\scriptsize F$\!$\raise-0.03ex\hbox{\it n}}}} }  
\def\pFp{p_{\raise-0.3ex\hbox{{\scriptsize F$\!$\raise-0.03ex\hbox{\it p}}}} }  
\def\pFe{p_{\raise-0.3ex\hbox{{\scriptsize F$\!$\raise-0.03ex\hbox{\it e}}}} }  
\def\pFl{p_{\raise-0.3ex\hbox{{\scriptsize F$\!$\raise-0.03ex\hbox{\it l}}}} }  
\def\m@th{\mathsurround=0pt }
\def\eqalign#1{\null\,\vcenter{\openup1\jot \m@th
   \ialign{\strut$\displaystyle{##}$&$\displaystyle{{}##}$\hfil
   \crcr#1\crcr}}\,}
\newcommand{\vp}{\mbox{\boldmath $p$}}         
\newcommand{\vS}{\mbox{\boldmath $S$}}
\newcommand{\vP}{\mbox{\boldmath $P$}}
\newcommand{\om}{\mbox{$\omega$}}              
\newcommand{\Om}{\mbox{$\Omega$}}              
\newcommand{\Th}{\mbox{$\Theta$}}              
\newcommand{\ph}{\mbox{$\varphi$}}             
\newcommand{\del}{\mbox{$\delta$}}             
\newcommand{\Del}{\mbox{$\Delta$}}             
\newcommand{\lam}{\mbox{$\lambda$}}            
\newcommand{\Lam}{\mbox{$\Lambda$}}            
\newcommand{\ep}{\mbox{$\varepsilon$}}         
\newcommand{\ka}{\mbox{$\kappa$}}              
\newcommand{\dd}{\mbox{d}}                     
\newcommand{\vect}[1]{\bf #1}                
\newcommand{\vtr}[1]{\mbox{\boldmath $#1$}}  
\newcommand{\vF}{\mbox{$v_{\mbox{\raisebox{-0.3ex}{\scriptsize F}}}$}}  
\newcommand{\pF}{\mbox{$p_{\mbox{\raisebox{-0.3ex}{\scriptsize F}}}$}}  
\newcommand{\kF}{\mbox{$k_{\rm F}$}}           
\newcommand{\kTF}{\mbox{$k_{\rm TF}$}}         
\newcommand{\kB}{\mbox{$k_{\rm B}$}}           
\newcommand{\tn}{\mbox{$T_{{\rm c}n}$}}        
\newcommand{\tp}{\mbox{$T_{{\rm c}p}$}}        
\newcommand{\te}{\mbox{$T_{eff}$}}             
\newcommand{\ex}{\mbox{\rm e}}                 
\newcommand{\rate}{\mbox{$\frac{ \mbox{▄╥╟}}{\mbox{╙═$^3 \cdot $╙}}$}}
\newcommand{\mur}{\raisebox{0.2ex}{\mbox{\scriptsize (э)}}} 
\newcommand{\Mn}{\raisebox{0.2ex}{\mbox{\scriptsize (э{\it n\/})}}}        %
\newcommand{\Mp}{\raisebox{0.2ex}{\mbox{\scriptsize (э{\it p\/})}}}        %
\newcommand{\MN}{\raisebox{0.2ex}{\mbox{\scriptsize (э{\it N\/})}}}        %
\begin{abstract}
We consider cooling of neutron stars (NSs) with superfluid
cores composed of neutrons, protons, and electrons
(assuming singlet-state pairing of protons, and triplet-state
pairing of neutrons).
We mainly focus on (nonstandard)
triplet-state pairing of neutrons with the
$|m_J| = 2$ projection of the total angular
momentum of Cooper pairs onto quantization axis.
The specific feature of this pairing is that it leads to
a power-law (nonexponential) reduction of the
emissivity of the main
neutrino processes by neutron superfluidity.
For a wide range of neutron critical temperatures $T_{cn}$,
the cooling of NSs
with the $|m_J| = 2$ superfluidity is either the same as
the cooling with the $m_J = 0$ superfluidity,
considered in the majority of papers,
or much faster.
The cooling of NSs with density dependent critical temperatures
$T_{cn}(\rho)$ and $T_{cp}(\rho)$ can be imitated by
the cooling of the NSs
with some effective critical temperatures $T_{cn}$
and $T_{cp}$ constant over NS cores.
The hypothesis of strong
neutron superfluidity with $|m_J| = 2$
is inconsistent with current observations of thermal
emission from NSs,
but the hypothesis of
weak neutron superfluidity of any type
does not contradict to observations.
\end{abstract}

\section{Introduction}
The cooling of neutron stars (NSs) strongly depends on
the properties of the matter in their cores, primarily on the
equation of state for the matter (Lattimer and Prakash
2001) and on the critical temperatures of NS nucleon
superfluidity (Lombardo and Schulze 2001). At present,
both are known incompletely, because there is no reliable
microscopic theory. The available knowledge about the
properties of the matter in NS cores can be improved by
comparing observational data with NS cooling models.

Here, we consider some cooling models. For definiteness,
we assume the standard composition of NS cores:
neutrons (n), protons (p), and electrons (e).

It is well known that the neutrons and protons in NS
cores can be in a superfluid state (Yakovlev et al. 1999;
Lombardo and Schulze 2001). Model calculations of
superfluid gaps show that proton pairing takes place in the
singlet $^1{\rm S}_0$ state of the proton pair. Neutron pairing can
take place in the singlet and triplet ($^3{\rm P}_2$) states; singlet
pairing generally arises in matter of moderate density
( $\rho \la \rho_0$,
where $\rho_0 = 2.8 \times 10^{14}$ g~cm$^{-3}$
is the nuclear density) and
triplet pairing arises in denser matter. The triplet pairing, in
turn, can be of several types that differ by the component of
the total moment of the nucleon pair along the quantization
axis ($|m_J| = 0, 1, 2)$. Following the
review articles by Yakovlev et al. (1999, 2001), we call the
singlet pairing case $A$, the triplet pairing with $|m_J| = 0$ case
$B$, and the triplet pairing with $|m_J| = 2$ case $C$. Case $C$
stands out among the remaining cases in that the superfluid
gap in the dispersion relation for neutrons becomes zero at
some points of the Fermi surface (at its poles). This leads to
a fundamentally different (power-law rather than
exponential) dependence of the NS neutrino energy losses on
the gap amplitude (Yakovlev et al. 1999). The type of
superfluidity with a minimum free energy occurs in nature.
The $A$-type proton pairing in the NS core, the $A$-type
neutron pairing in the crust and in the outer core, and the
$B$-type pairing in the inner core are commonly considered
in cooling calculations. The $C$-type neutron pairing in the
NS core seems less realistic, but it is not ruled out by
current microscopic theories.

Up until now, virtually nobody has modeled the NS
cooling with the $C$-type neutron superfluidity. We know
only the recent paper by Schaab et al. (1998), who
attempted to take into account the effect of such pairing on
the main processes of neutrino energy release in NS cores
and on the NS cooling. Unfortunately, the authors failed to
completely consider all of the factors that determine the
thermal evolution of NSs. Thus, for example, when
considering one of the most important neutrino processes,
the neutrino emission during Cooper neutron pairing, the
authors used an exponential dependence instead of the
correct power-law dependence of the rate of energy release on
$T/T_{cn}$ ($T_{cn}$ is the critical neutron temperature).

Here, we analyze the NS cooling for the $C$-type neutron
pairing more rigorously. We compare the coolings for the
$B$- and $C$-type superfluidities. Our results are compared
with observations.
\section{NS cooling for the $C$-type neutron pairing}
For the cooling to be adequately modeled, we must
know the rate of neutrino energy release in superfluid NS
interiors and the heat capacity of the NS matter. Recall that
the main neutrino processes in NS cores are the direct Urca
process, the modified Urca process, the nucleon-nucleon
scattering reactions, and the Cooper pairing of nucleons.
The first process, the most powerful mechanism of neutrino
energy release, is a threshold process. The direct Urca
process is open if $p_{{\rm F}n} \leq 2 p_{{\rm F}p}$,
where $p_{{\rm F}n}$ and $p_{{\rm F}p}$ are the
neutron and proton Fermi momenta, respectively.

The neutrino processes and heat capacity for cases $A$, $B$,
and $C$ have been studied extensively. The results can be
found in the review article by Yakovlev et al. (2001) and in
Gusakov (2002). Yakovlev et al. (2001) considered the
neutrino processes with the $A$- or $B$-type neutron and
A-type proton superfluidities;
they took into account the effect of the $C$-type superfluidity
on the direct Urca process, the neutrino energy release
during Cooper neutron pairing, and the neutron heat
capacity. Gusakov (2002) investigated the modified Urca
process during the $C$-type pairing and the $nn$- and
$np$-scattering reactions. He also rigorously took into account
the effect of the combined $A$- or $B$-type nucleon
superfluidity on the modified Urca process.

Here, we use the relativistic nonisothermal NS cooling
code (see Yakovlev et al. 2001) adapted for the model with
the $C$-type neutron superfluidity. This code allows us to
construct the cooling curves, i.e., to determine the
dependence of the surface temperature of a star, $T_e^{\infty}$,
(recorded by a remote observer with allowance made for
the gravitational redshift) on its age, $t$. Following Kaminker
et al. (2002), we use the equation of state by Negele and
Vautherin (1973) in the NS crust and the equation of state
by Prakash et al. (1988) in the NS core with the
compression modulus $K = 240$ MeV for symmetric nuclear
matter at $\rho = \rho_0$
(model I from their paper for the symmetry
energy). For this equation of state, the direct Urca process
opens at $\rho \geq \rho_D = 7.851 \times
10^{14}$ g~cm$^{-3}$, i.e., at  $M \geq 1.358 M_{\odot}$. In
this case, the maximum NS mass is  $M = 1.977 M_{\odot}$. In the
cooling code used, the effective proton and neutron masses
in the inner NS layers renormalized with allowance made
for multiparticle effects were assumed to be
0.7 of the mass of the free particles. The relationship
between the NS surface and internal temperatures was
taken from Potekhin et al.~(1997).

To better understand at which critical nucleon
temperatures the $C$-type neutron superfluidity in NS cores
differs from case $B$, we mentally replace the NS under
consideration by a sphere with a radially constant density
characteristic of this star. In that case, after the completion
of thermal relaxation and before the beginning of the
photon NS cooling stage, the time derivative of the internal
NS temperature can be estimated as $dT/dt \sim -Q/C$, where
$Q$ is the energy released in the form of neutrinos per unit
volume per unit time and $C$ is the heat capacity per unit
volume. This formula holds for both $B$- and $C$-type neutron
superfluidities. Since the $C$-type superfluidity suppresses
the NS energy release more weakly and since the energy
released via Cooper pairing is higher than that in case $B$, we
may say that $Q_C > Q_B$ always. Similarly, $C_C > C_B$,
therefore, it is not clear in advance which stars cool down
faster: with the $C$- or $B$-type neutron superfluidity in the NS
core. If we introduce the critical nucleon temperatures $T_{cn}$
and $T_{cp}$ for the characteristic density under consideration,
then, depending on $T_{cn}$ and $T_{cp}$, the function
$\log[Q_C C_B/(Q_B C_C)]$ will be convenient for understanding
when the cooling for cases $C$ and $B$ proceeds almost
identically and when the differences are at a maximum.

Let us examine Fig. 1. 
\begin{figure*}[]
\begin{center}
\epsfysize=200mm
\epsffile[157 161 524 699]{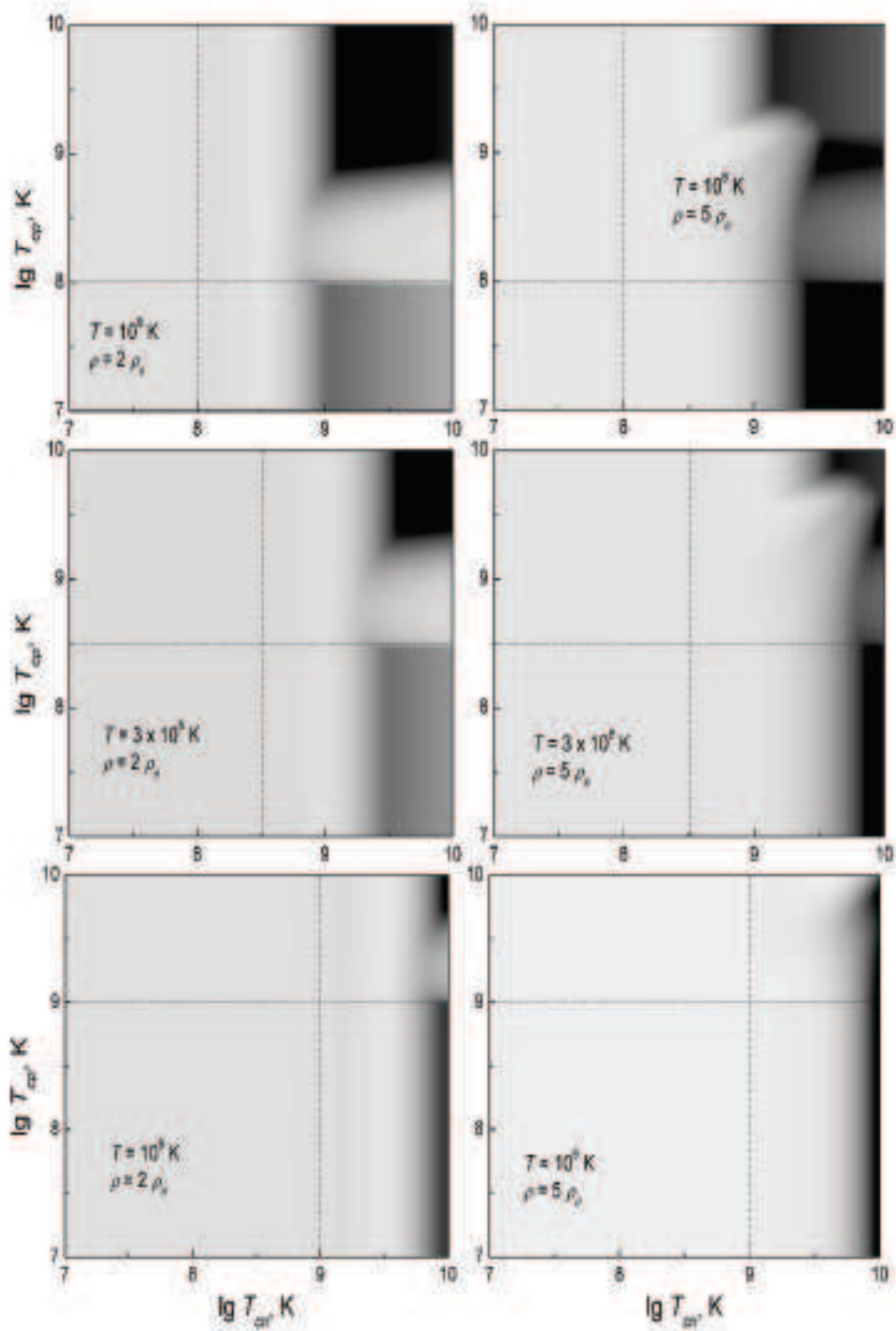}
\caption{The function $\log[Q_C C_B/(Q_B C_C)]$
over a wide range of $T_{cn}$ and $T_{cp}$
is indicated by different shades of gray. The larger is the
value of the function, the darker is the shade (see the text).
}
\end{center}
\label{Bcase}
\end{figure*}
It consists of six pictures, each
corresponding to a certain temperature ($T = 10^8$, $3 \times 10^8$, or
$10^9$ K) and density ($\rho = 2 \rho_0$, or $5 \rho_0$).
The direct Urca
process is forbidden at $\rho = 2 \rho_0$ and permitted at
$\rho = 5 \rho_0$.
Our calculations show that when the temperatures and
densities are varied, the pictures do not change
fundamentally until the threshold density $\rho_D$ at which the
direct Urca process switches on (or off) is crossed. The
logarithms of the critical proton and neutron temperatures
at a fixed density are plotted along the vertical and
horizontal axes of each picture, respectively. The critical
nucleon temperatures lower than $10^7$ K (weak superfluidity)
have virtually no effect on the NS cooling and, hence, are
not considered. In the figure, the function
$\log[Q_C C_B/(Q_B C_C)]$ is indicated by different shades of gray.
The shade of gray for the region of critical temperatures to
the left from the vertical dashed line (nonsuperfluid
neutrons) corresponds to zero of this function. The darker is
the shade of gray, the larger is $\log[Q_C C_B/(Q_B C_C)]$.
For example, for $T = 10^8$ K, the maximum of this function
reaches (6 -- 7) for critical proton and neutron temperatures
$T_{cp,cn} \sim (10^9-10^{10})$ K
at $\rho = 2 \rho_0$ and (8 -- 9) for $T_{cp} \sim (10^7-10^8)$
K, $T_{cn} \sim 3 \times (10^9 -10^{10})$ K at
$\rho = 5 \rho_0$. Conversely, the
lighter is the shade of gray
relative to the 'zero' shade, the smaller is the function. The
region with the shade of gray lighter than the 'zero' shade is
narrow and lies near $T_{cn} \sim 3 T$. In this region, the
minimum value of the function is of the order of --1 for
both densities. A further increase in the internal NS
temperature causes the differences between the $B$- and
$C$-type neutron superfluidities to be smoothed out; the
smoothing occurs faster for the densities at which the direct
Urca process is open. Thus, for $T = 10^9$ K, the maximum of
$\log[Q_C C_B/(Q_B C_C)]$ is (3 -- 3.5) for $\rho = 2 \rho_0$ and (1 -- 1.2)
for $\rho = 5 \rho_0$, while its minimum is nearly zero for both
densities. Summing up our results, we may say that the
inequality $(dT/dt)_C \geq (dT/dt)_B$ holds for any $T$, 
$T_{cn}$, $T_{cp}$,
and $\rho$. Therefore, one can hardly expect the NS cooling
curve for case $C$ to pass well above the cooling curve for
case $B$ (at fixed $T_{cn}$ and $T_{cp}$).
Clearly, the reverse can be easily realized by an
appropriate (see above) choice of critical temperatures.

Let us now turn directly to an examination of the NS
cooling curves. For simplicity, the neutrons in the stellar
crust are assumed to be nonsuperfluid. This assumption has
no effect on the difference between the cooling curves for
the $B$- and $C$-type neutron superfluidities in the NS core.
Although the critical nucleon temperatures $T_{cn}(\rho)$ and
$T_{cp}(\rho)$ are actually functions of the density and vary along
the stellar radius, it was noticed that following certain
semiempirical recipes, constant (over the core) critical
temperatures can be matched to them. In this case, the cooling
curves remain the same as if we took into account the exact
dependence $T_{cn}(\rho)$ and $T_{cp}(\rho)$
(e.g., for protons, a constant
effective temperature $T_{cp}$ close
to $T_{cp}(\rho_c)$, where $\rho_c$
is the central density of the star,
always corresponds to the temperature profile $T_{cp}(\rho)$).
Therefore, for simplicity, we assume the critical nucleon
temperatures to be density-independent.
\begin{figure}[t]
\begin{center}
\epsfysize=80mm
\epsffile[71 212 553 678]{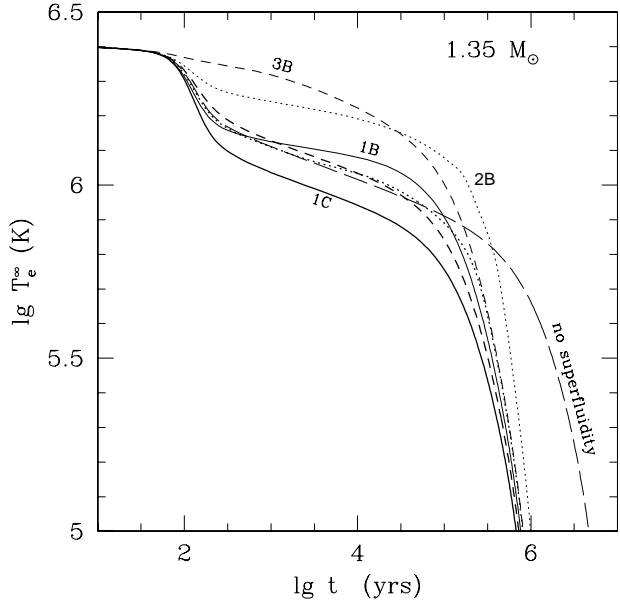}
\caption{
The cooling curves constructed for
the $B$- and $C$-type neutron superfluidities
in NS cores for various critical nucleon
temperatures. The curves with identical
$T_{cn}$ and $T_{cp}$ but for different
superfluidities, $B$ or $C$, are represented by the same type of
lines (solid, dotted, and dashed lines).
The long dashes correspond to the
cooling without superfluidity. The NS mass is
$M = 1.35 M_{\odot}$.}
\end{center}
\label{1}
\end{figure}
\begin{figure}[t]
\begin{center}
\epsfysize=80mm
\epsffile[71 212 553 678]{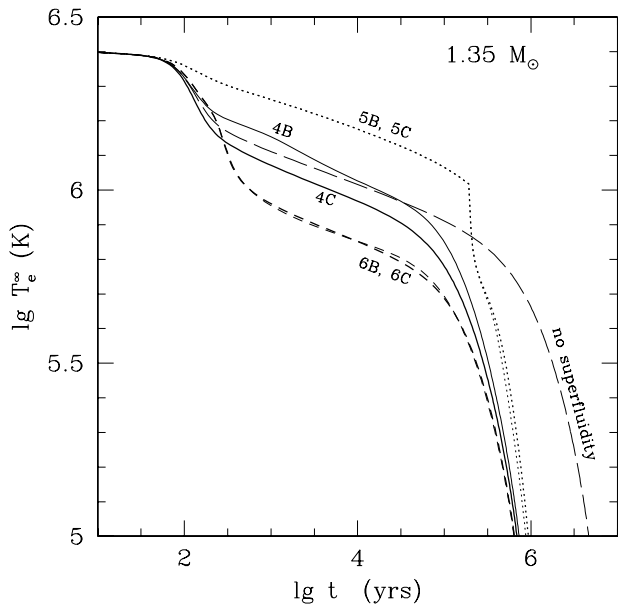}
\caption{Same as Fig. 2 but for another superfluidity models}
\end{center}
\label{2}
\end{figure}
\begin{figure}[t]
\begin{center}
\epsfysize=80mm
\epsffile[71 212 553 678]{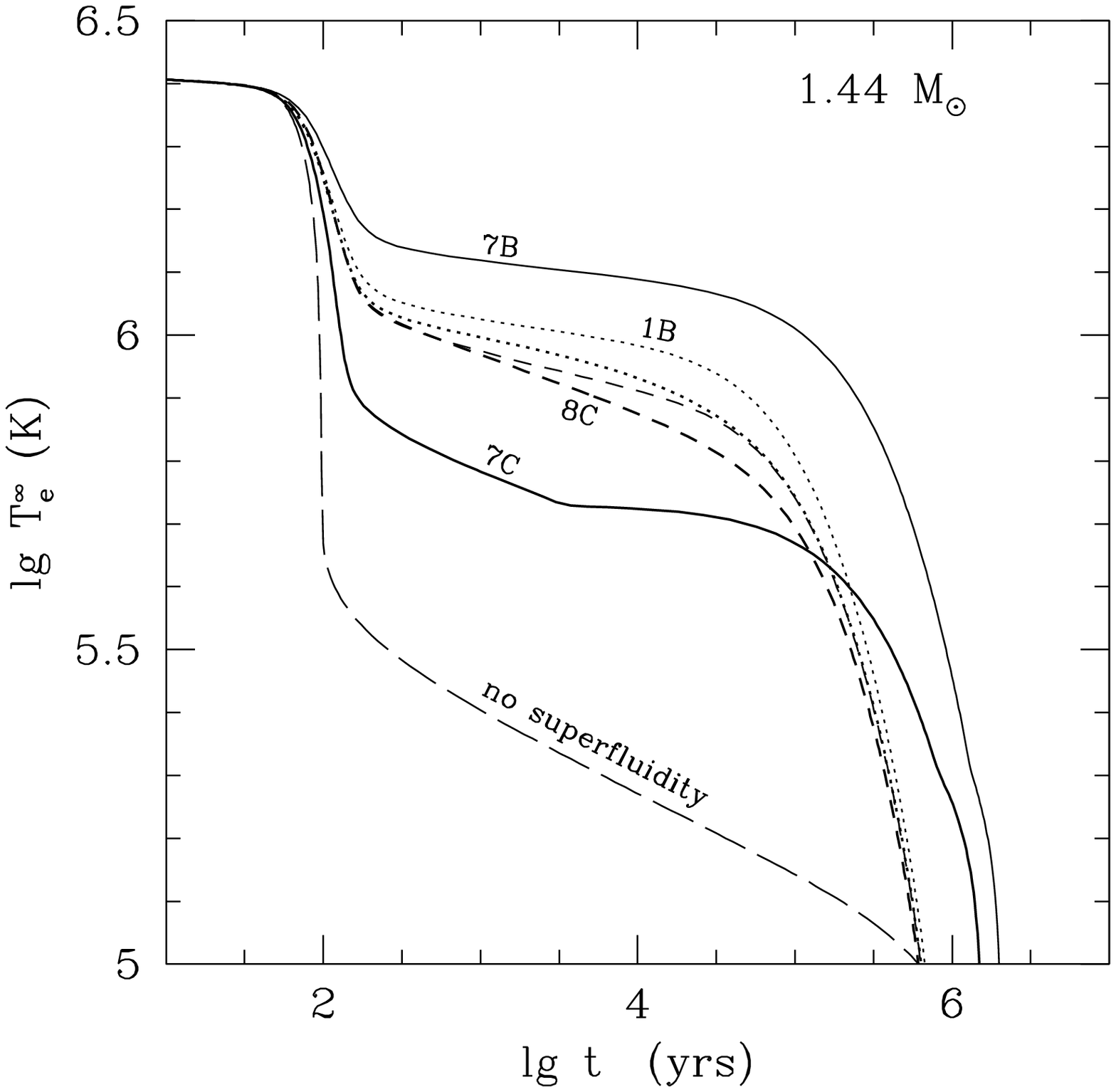}
\caption{ Same as Figs. 2 and 3 for $M = 1.44 M_{\odot}$}
\end{center}
\label{3}
\end{figure}
\begin{figure}[t]
\begin{center}
\epsfysize=80mm
\epsffile[71 212 553 678]{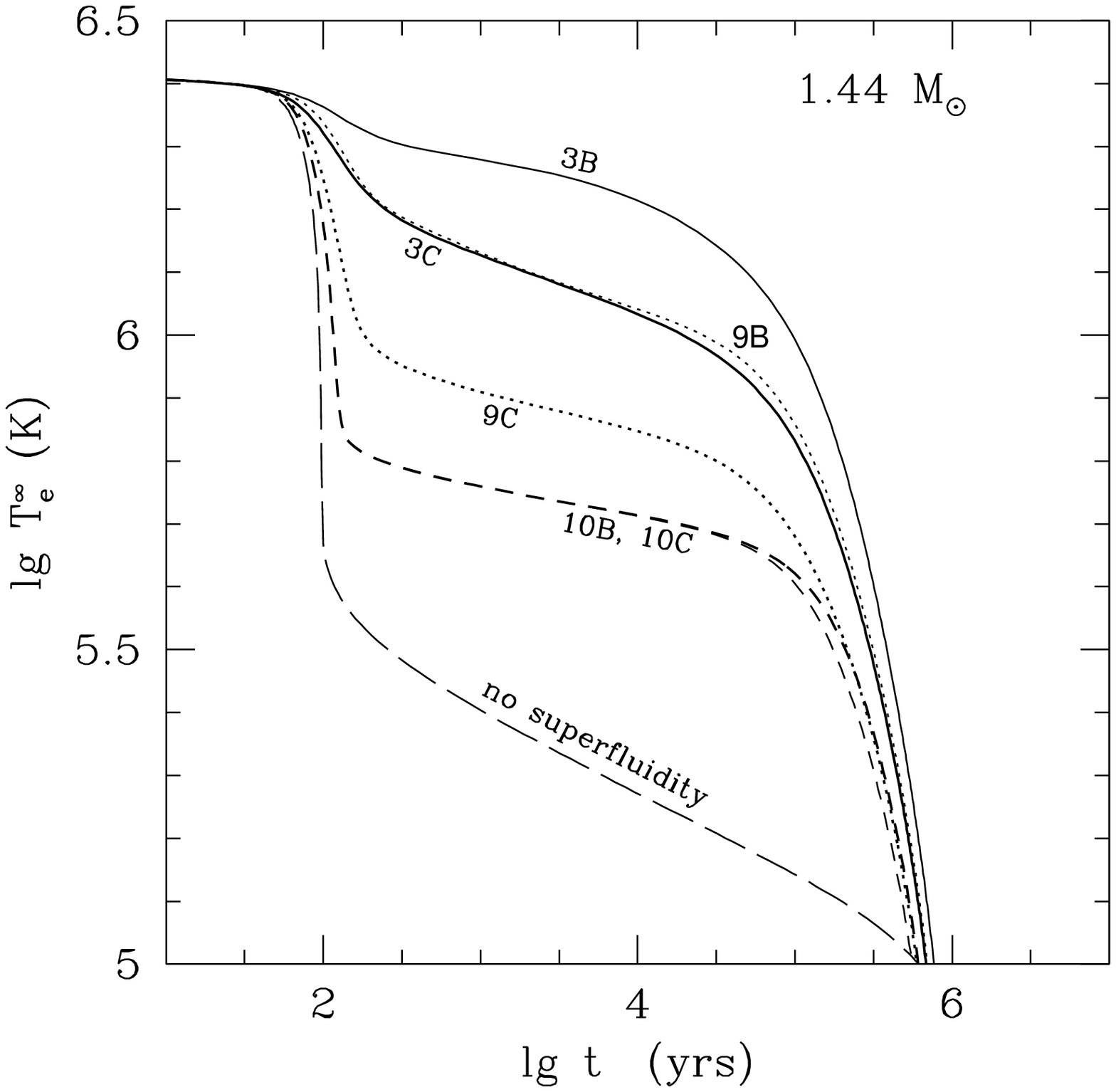}
\caption{Same as Figs. 2 and 3 for $M = 1.44 M_{\odot}$}
\end{center}
\label{4}
\end{figure}
\begin{figure}[t]
\begin{center}
\epsfysize=80mm
\epsffile[71 212 553 678]{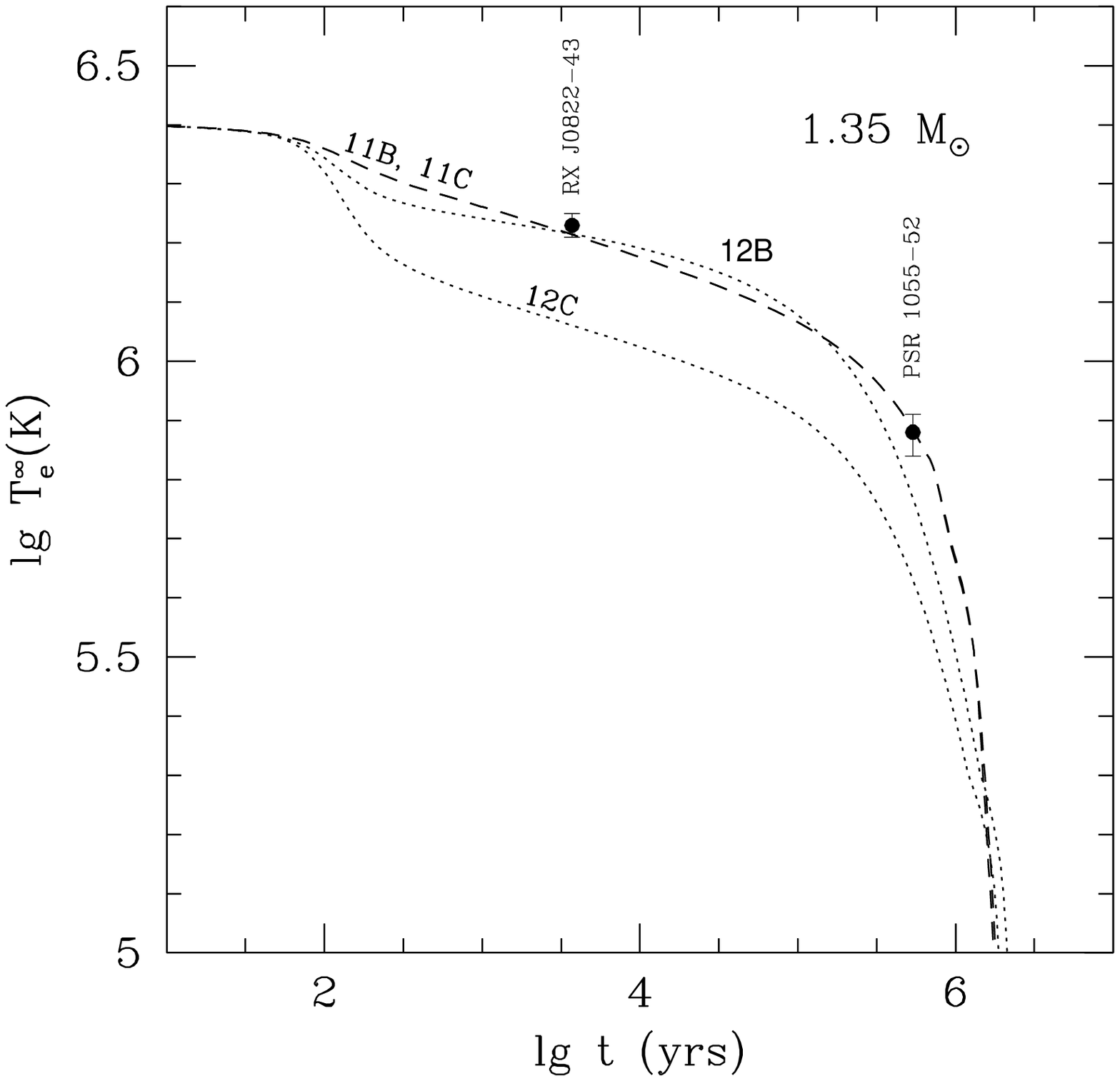}
\caption{The cooling curves for the models with the $B$- and
$C$-type neutron superfluidities in the core of a NS with
$M = 1.35 M_{\odot}$ and observational data on the surface temperatures
of two sources. The dashed curves are for $T_{cn} = 10^8$ K and
$T_{cp} = 5 \times 10^9$ K;
the dotted curves are for $T_{cn} = 5 \times 10^9$~K and
$T_{cp} = 10^7$~K.
}
\end{center}
\label{6}
\end{figure}

Figures 2 -- 5 show the cooling curves constructed for
two NS masses:  $M = 1.35 M_{\odot}$ (Figs. 2 and 3)
and $M = 1.44 M_{\odot}$ (Figs. 4 and 5). The direct Urca process is
forbidden for  $M = 1.35 M_{\odot}$ and permitted for
$M = 1.44 M_{\odot}$. In Table 1
we give the various critical nucleon temperatures $T_{cn}$ and
$T_{cp}$ for which the cooling curves were constructed.
\begin{table}[t]
\caption{
Chosen nucleon superfluidity models.
}
\begin{tabular}{|c|r|r|r|r|r|r|r|r|r|r|r|r|}
\hline
Number&1&2&3&4&5&6&7&8&9&10&11&12 \\
\hline
$T_{cn}/10^9$, K&2.5&5.0&10.0&3.2&0.2&0.5&7.0&0.85&9.5&0.2&0.1&5.0 \\
\hline
$T_{cp}/10^9$, K&2.5&0.25&10.0&0.6&5.0&1.0&0.01&3.0&0.85&0.7&5.0&0.01 \\
\hline
\end{tabular}
\end{table}
Each pair of critical temperatures in the table has its own
number. In Figs. 2 -- 6, the cooling curves are marked by
these numbers. The letter $B$ or $C$ near the number specifies
the type of neutron superfluidity.
For example, $3C$ means that the cooling curve was
constructed for the $C$-type neutron superfluidity for the
critical temperatures $T_{cn}$ and $T_{cp}$ under number 3. We see
from analysis of Figs. 2 -- 5 that all
our main conclusions drawn from an examination of Fig. 1
are confirmed. For $M = 1.35 M_{\odot}$, the largest differences are
observed for the $3B$ and $3C$ curves in Fig. 2 (high critical
temperatures). At critical proton temperatures
$T_{cp} \sim 5 \times 10^8$ K,
the neutrino emission during Cooper proton pairing is
the most powerful process in the NS core. Accordingly, the
differences between the cooling curves for cases $B$ and $C$
decrease. This can be verified by comparing the $4B$ and $4C$
curves in Fig. 3. As the critical proton temperatures
decrease to $T_{cp} \la 3 \times 10^8$ K and for $T_{cn} \ga 10^9$ K
(but $T_{cn} \la 10^{10}$~K),
the differences again increase (the $2B$ and $2C$
curves in Fig. 2). When $T_{cn} \ga 10^{10}$~K and increases, the
differences gradually decrease and the cooling curves for
the $B$- and $C$-type neutron paring coincide
for $T_{cn} \sim 10^{11}$ K.
At moderately high critical temperatures $T_{cn} < 10^9$ K, the
curves for case $B$ almost coincide with those for case $C$
(see, e.g., Fig. 3: $5B$ and $5C$ or $6B$ and $6C$). The kink in the
$5B$ and $5C$ curves occurs at $T = T_{cn}$ and stems from the fact
that the neutrino emission due to Cooper neutron pairing
switches on. Note that the critical temperatures of the $6B$
and $6C$ curves exactly fall within the region where the
temperature for the $B$-type neutron superfluidity decreases
faster than that for case $C$ ($\log[Q_C C_B/(Q_B C_C)] < 0$).
This actually leads to the fact that the $6B$ curve initially passes
slightly below the $6C$ curve. Subsequently, as the internal
temperature $T$ decreases, the situation is reversed, because
we reach the regions with faster cooling for the $C$-type
superfluidity (see Fig. 1).

For $M = 1.44 M_{\odot}$, the differences are largest for the $7B$ and
$7C$ curves in Fig. 4, i.e., at high critical neutron
temperatures and at low critical proton temperatures. As
$T_{cn}$ decreases, the differences between the cooling curves
for the $B$- and $C$-type superfluidities gradually disappear
(see Fig. 4: $8B$ and $8C$; Fig. 5: $10B$ and $10C$). The
remaining remarks almost completely repeat those for
$M = 1.35 M_{\odot}$.

Our results can be of use in explaining observational
data. The NS cooling with weak $B$-type neutron
superfluidity was considered by Kaminker et al. (2002).
The authors showed that the available observational data on
the thermal radiation from eight isolated NSs could be
explained by assuming
strong proton ($T_{cp,{\rm max}} \ga 5 \times 10^9$ K) and weak neutron
($T_{cn,{\rm max}} \la 10^8$ K) superfluidities.
In this case, the critical
temperatures $T_{cp}$ and $T_{cn}$ must depend on density $\rho$
[have a maximum at $\rho \sim (2-3) \rho_0$,
as predicted by microscopic
theories of superfluidity; see, e.g., Lombardo and 
Schulze~(2001)]. The mechanisms of strong proton and weak
neutron superfluidities are primarily needed to interpret the
observations of two sources, RX J0822--43 (Zavlin et al.
1999) and PSR 1055--52 ($\ddot {\rm O}$gelman 1995), the hottest ones
for their age. The observational data on these sources are
shown in Fig. 6 [we took the same $T_e^{\infty}$ and $t$ as those used
by Kaminker et al. (2002)]. Until recently, in addition to
these two sources, yet another source, RX J185635--3754,
must have been considered (Pons et al. 2002; Burwitz et al.
2001; G$\ddot {\rm a}$nsicke et al. 2001; Kaplan et al. 2001). However,
in the just published paper by Walter and Lattimer (2002),
the age of RX J185635--3754 was revised (reduced by
almost half) and is now estimated to be $t \approx 5 \times 10^5$ yr,
which is attributable to a decrease in the distance to this
source.

When comparing observations with cooling models, we
should take into account the fact that the surface
temperatures and ages of isolated cooling NSs are
determined with a low accuracy [for the reasons discussed,
e.g., by Pavlov and Zavlin (1998) and Yakovlev et al.
(1999)]. The NS age $t$ is probably known to within a factor~$\sim~2$.

Since the determination of $T_e^{\infty}$ and $t$ is ambiguous, we
can offer an interpretation of the observations alternative to
that offered by Kaminker et al. (2002). More specifically, if
we take the lowest possible values of $T_e^{\infty}$ and $t$ for
PSR~1055--52, then the observations of the eight NSs considered
by Kaminker et al. (2002) can be interpreted by assuming
strong neutron superfluidity and weak proton superfluidity
in the stellar cores. Although this interpretation is less
plausible than that offered by Kaminker et al. (2002)
(according to microscopic calculations of $T_{cn}(\rho)$ and
$T_{cp}(\rho)$, the triplet neutron pairing is generally weaker than
the single proton pairing), it cannot be ruled out in advance
and should be studied. It actually requires lower values of
$T_e^{\infty}$ and $t$ of the fairly old and hot
source PSR~1055--52, because strong neutron superfluidity
significantly reduces the NS heat capacity and speeds up
the NS cooling at the late photon cooling stage [implying
that the models by Kaminker et al. (2002) can describe
older and hotter sources].

A detailed discussion of the hypothesis of strong neutron
superfluidity and weak proton superfluidity in NS cores is
beyond the scope of this paper and will be given in our
subsequent publication. Here, we only illustrate the
possibility of this interpretation of observations for the two
hottest sources (RX~J0822--43 and PSR~1055-52) by using
simplified models of cooling NSs with critical temperatures
$T_{cn}(\rho)$ and $T_{cp}(\rho)$
constant over the stellar core. The dotted
lines in Fig. 6 represent the cooling curves for $T_{cn} = 5 \times
10^9$ K and $T_{cp} = 10^7$ K. For comparison, the dashed line
was drawn for strong proton superfluidity and weak neutron
superfluidity ($T_{cp} = 5 \times 10^9$ K, $T_{cn} = 10^8$ K).
It virtually coincides with the cooling curve obtained by Kaminker et
al. (2002) to interpret the above two sources. We see that
the upper dotted line ($B$-type neutron superfluidity) almost
coincide with the dashed line for $t \la 10^5$ yr and passes
slightly below the dashed line for $t \ga 10^5$ yr (because of the
above suppression of the neutron heat capacity by strong
neutron superfluidity). Nevertheless, given the inaccurate
determination of $T_e^{\infty}$ and $t$,
the upper dotted line could be
assumed to satisfactorily explain these two sources. At the
same time, the lower dotted line for strong $C$-type neutron
superfluidity passes well below (because of the weaker
suppression of neutrino emission by the $C$-type
superfluidity) and cannot be considered to be acceptable.
Therefore, the observations can (with the above
reservations) be explained by strong $B$-type neutron
superfluidity but cannot be explained by the $C$-type neutron
superfluidity with the same $T_{cn} = 5 \times 10^9$ K. With a further
increase in $T_{cn}$, the upper dotted line would not change at
all (strong $B$-type neutron superfluidity with
$T_{cn} \sim 5 \times 10^9$ K
almost completely switched off the neutrino reactions
involving neutrons and made the neutron heat capacity
equal to zero), while the lower dotted line would slowly
approach the upper curve (the $C$-type neutron superfluidity
increasingly suppresses the neutrino luminosity and heat
capacity), reaching it at $T_{cn} \sim 10^{11}$ K.
Thus, the interpretation of the observations
in the model of strong $C$-type
neutron superfluidity is possible in principle but is
implausible, because it requires unrealistically high values
of $T_{cn} \sim 10^{11}$ K.
\section{Conclusions}
The cooling of NSs with $C$-type neutron superfluidity in
their cores has been consistently studied for
the first time. The critical nucleon temperatures were
assumed to be constant over the stellar core. A comparison
with $B$-type neutron superfluidity was made. We found that
the cooling curves for case $C$ can pass well below the
cooling curves for case $B$ (faster cooling). Otherwise, the
curves differ only slightly. The cooling with variable (over
the core) critical temperatures $T_{cn}(\rho)$ and $T_{cp}(\rho)$ can
generally be described by the cooling with some effective
constant temperatures $T_{cn}$ and $T_{cp}$.

The model with $C$-type neutron superfluidity can
account for the observational data on the thermal radiation
from isolated NSs by assuming strong proton superfluidity
($T_{cp} \ga 5 \times 10^9$ K)
and weak neutron superfluidity ($T_{cn} \la 10^8$ K).
The interpretation of the observations in terms of the
model with strong neutron superfluidity and with weak
proton superfluidity ($T_{cp} \la 10^8$ K)
is unlikely (in contrast to
the same model for case $B$), because it requires too high
critical neutron temperatures, $T_{cn} \sim 10^{11}$ K. Our results can
be useful in understanding which type of superfluidity ($B$ or
$C$) occurs in nature.
\section{Acknowledgments}
One of the authors (MEG) is grateful to D.G.~Yakovlev and A.D.~Kaminker
for helpful discussions and attention to the work.
This study was supported in part by the Russian Foundation
for Basic Research (project nos. 02-02-17668 and 00-07-90183).

\end{document}